\documentstyle[11pt]{article}
\begin{document}
\bibliographystyle{unsrt}
\def\f {\beta_f}
\def\a {\beta_a}
\def\l {\alpha}

\setcounter{page}{0}
\thispagestyle{empty}

\vskip3cm
\begin{flushright}
TIFR/TH/96-55 \\
hep-lat/9610022
\end{flushright}

\vskip2cm
\begin{center}
{\Large {\bf Stability of the Bulk Phase Diagram \\
of the SU(2) Lattice Gauge Theory \\
with Fundamental - Adjoint Action}}\\[1cm]

{\large Saumen Datta \footnote{E-mail:saumen@theory.tifr.res.in} and
Rajiv V. Gavai \footnote{E-mail:gavai@mayur.tifr.res.in}}\\[1cm]

{\em Theoretical Physics Group, \\ Tata Institute of Fundamental
Research, \\
Homi Bhabha Road, Mumbai - 400005, India.}\\[2cm]

{\large\bf Abstract}
\end{center}

Using improved mean field and strong coupling expansions we re-analyse
the bulk phase diagram of the fundamental-adjoint action of the SU(2)
Lattice Gauge Theory. We find that the qualitative features of the
bulk phase diagram are robust and unchanged by the inclusion of higher
order terms. On the other hand, some of the quantitative features, such
as the location of the endpoint of the line of bulk phase transitions,
seem to be strongly dependent on the higher order terms of the strong
coupling expansion.

\newpage

Lattice regularization of a continuum action is not unique.
For non-abelian gauge theories the Wilson action \cite{wil} is the
most popular one, but other actions have been studied in the literature.
In particular, lately there has been a resurgence of
interest in the 
Bhanot-Creutz action\cite{bha}
\begin{equation}
S=\sum_p \Bigl\lbrace \f \bigl (1 - {1 \over 2} Tr_F U_p \bigr )+ \a
\bigl ( 1 - {1 \over 3} Tr_A
U_p \bigr ) \Bigr\rbrace. \label{BHA}
\end{equation}
Here F and A denote the fundamental and
adjoint representations respectively.
The Wilson action is a special case of (\ref{BHA}), corresponding to $\a = 0$.
The action (\ref{BHA}) was first studied by Bhanot and Creutz for SU(2) gauge theories in order to understand
the bulk phase transition found in numerical studies of some
non-abelian gauge theories (SO(3), SU(4) etc.) and the role they play
in the physics of confinement.
They found a line of first order transition in the $\f$ - $\a $ plane
(see Fig.1) that ended
at a finite $\a$. Since the location of the peak in the plaquette
susceptibility 
for the Wilson action corresponds to the interception of the extrapolation of this line
with the $\a = 0$ axis, it has been considered as a possible source
for the observed crossover in the string tension. It is thus a
possibly important key in our understanding of the confinement phenomenon.

However, recent \cite{gav1,mat} finite temperature
investigations of this action have cast some doubt on the nature of
the phase transition line seen in \cite{bha}.
It was found that switching on a nonzero $\a$, the known finite temperature
phase transition of the Wilson action becomes a line and joins the
above mentioned bulk transition line. Moreover, the order of the deconfinement
transition  changes
from second to first order at $\a \ge 1.5$. No indications of two
separate transitions were found at any $\a$.
After considering various possibilities, it was concluded \cite{mat}that the
transition line is not a bulk one, but the
deconfinement transition line. Since the study in
\cite{bha} was done on 
small lattices, which were at relatively high temperature, such a
misinterpretation is possible.

A finite size scaling-based
analysis of the bulk transition has also been done \cite{gav3} to
determine the nature of the phase transition. It was found that the line
of the phase transition ends at a somewhat higher $\a$ than what was found
in \cite{bha}. Simulations at a still higher value of $\a$ (=1.5)
suggested the presence of a 1st order bulk phase transition, but the Polyakov line, which is the order parameter of the
finite temperature phase transition, was also found to jump across this
transition at $\a = 1.5$, thus further making a distinction between
the zero temperature bulk transition and the finite temperature
deconfinement transition very difficult.

Numerical simulations thus seem to give conflicting signals. The deconfinement order parameter acquires non-zero vacuum
expectation value at these transitions for all $\a$. However, the
shift of the transition point with $N_t$ - the temporal lattice size - 
decreases and becomes negligibly small. The
latter is suggestive of a bulk transition unless this behaviour
changes on very large lattices. In the absence of such large lattice simulations it
may be instructive to look for guidance by conventional analytical
techniques.

The existence of a deconfinement phase transition for $\a = 0$ has been
rigorously proven \cite{bor} in the strong coupling limit.
In \cite{gav2} a leading order strong coupling analysis of the action in
(\ref{BHA}) was done at finite temperature. It yielded a deconfinement phase transition
line in the $(\f, \a)$ plane, along with a change in order of the phase
transition for large $\a$, as seen in \cite{gav1}.

 A mean-field analysis of (\ref{BHA}) at zero temperature was done in
Ref \cite{alb} and the results of \cite{bha} were supported. In view
of the results of
\cite{gav3}, however, it seems to be necessary to reexamine the
results of Ref \cite{alb}. In this note, we have attempted to check how stable the
results of Ref \cite{alb} are, both qualitatively and quantitatively, by
improving and extending their work to higher orders in a consistent
manner. We find that our results still predict qualitatively the same
bulk phase diagram as in \cite{alb} but the location of the endpoint of the transition
line is very sensitive to higher orders and
cannot be precisely obtained by our study.

The mean field analysis proceeds by writing the partition function 
\begin{equation}
Z(\f, \a) = \int \prod_l dU_l \exp (- S)
\end{equation}
in the axial gauge by fixing all the links in time direction equal to 1.
The SU(2) elements are parametrized as $U = u^o + i {\bf u}.{\bf
\sigma}$, for real numbers $(u_0, {\bf u}) $ satisfying $u_0^2 +
u_1^2 + u_2^2 + u_3^2 = 1$. The standard technique \cite{bre, dro} of
Fourier transforming the measure and the action gives
\begin{equation}
Z(\f, \a) = \int \prod_l dv_l \int \prod_l {d\l_l \over (2 \pi i)^4}
\exp \bigl ( - S(\f, \a, v_l)  + \sum_l \bigl(w(\l) - \l^o_l
v^o_l - {\bf \l}_l . {\bf v}_l\bigr) \bigr ) \label{ACTION}
\end{equation}
where
\begin{equation}
w(\l)=\ln \int d\Omega \exp(\l^o u^o + {\bf \l}.{\bf u}).
\end{equation}

One then looks for translationally invariant saddle points of (\ref{ACTION}) of the form \cite{bre}
\begin{equation}
v_l = (v, {\bf 0}), \l_l = (\l, {\bf 0}).
\end{equation}

For $\f < 2$ one gets the stable solution $v=0,\l=0$ \cite{alb}. An
expansion of the free energy around this saddle point gives just the strong coupling expansion. The strong coupling expansion for the Bhanot-Creutz action has been obtained upto terms of order sixteen by Dashen et al \cite{das}. In our notation the series for free energy as obtained by them is
\begin{eqnarray}
F_I(\f, \a) = &{}& 6 \ln b_0 + 4 \bigl(4 b_1^6 + 9 b_2^6\bigr) + 36 \bigl(4 b_1^{10} + 9
b_2^{10}\bigr) \nonumber\\&{}& + 36 \bigl(12 b_1^{10} b_2 + 24 b_1^6 b_2^5 + 27
b_2^{11} - 224 b_1^{12} - 1368 b_1^6 b_2^6 - 1359 b_2^{12}\bigr) \nonumber \\&{}&
+ 4 \bigl(432 b_1^5 b_2^5 b_3 + 405 b_2^{10} b_4 + 16 b_3^6 + 25 b_4^6\bigr)
\label{STRONG}
\end{eqnarray}
and the characters $b_0$ and $b_j$, j=1,2,3,4, are evaluated as
\begin{eqnarray}
b_0 & = & \int_0^{4 \pi} {d\theta \over 2 \pi} \sin^2{\theta \over 2}
\exp \Bigl( -\f \bigl ( 1 - \cos{\theta \over 2} \bigr ) - {2 \over 3}
\a (1 - \cos \theta) \Bigr), \\
b_j & = & {1 \over (j+1) b_0} \int_0^{4 \pi} {d\theta \over 2 \pi}
\sin{\theta \over 2} \sin{(j+1) \theta \over 2} 
\exp \Bigl( -\f \bigl ( 1 - \cos{\theta \over 2} \bigr ) - {2 \over 3}
\a (1 - \cos \theta) \Bigr).
\end{eqnarray}
The characters are now expanded in a series of $\f$ and $\a$ to rewrite
$F_I$ explicitly as an expansion in powers of $\f$, $\a$. We checked that for our region
of interest, the difference between eqn(\ref{STRONG}) and the explicit
series is negligible. Also the convergence of the strong coupling
series was checked by looking at the series terminated at differet
orders. It was found that the strong coupling series converges very
slowly near the transition line, specially for smaller $\a$ values. 

For regions of large $\a$ and $\f$ there are stable
solutions of (\ref{ACTION}) for nonzero $v, \l$ satisfying the
equations \cite{alb}
\begin{eqnarray}
v & = & {I_2 (\l) \over I_1 (\l)}, \\
\l & = & \f 2 v \bigl(1+(d-2) v^2\bigr) + \a {16 \over 3} v^3 \bigl(1+(d-2) v^4\bigr),
\end{eqnarray}
where $I_2$, $I_1$ are the modified Bessel functions; d, the space-time dimensionality, will be 4 in all our
calculations.

Expanding near this saddle point, the partition function, or equivalently, the free energy per site, is
\begin{equation}
F_{II}(\f, \a) = {1 \over N} \ln Z(\f, \a) = F_{tree} + F_{1-loop} +
F_{2-loop} + ... \label{FOUR}
\end{equation}
The first two terms of this series were calculated in Ref \cite{alb} and are
\begin{eqnarray}
F_{tree} & = & 3 \f \bigl(v^2 + v^4 - 2 \bigr ) + 4 \a (v^4 + v^8 - 2
\bigr ) + 3 \Bigl(\ln{2 I_1(\l) \over \l} - \l v \Bigr), \\
F_{1-loop} & = & {3 \over 2} \Bigl(\ln 2+3 \ln {\l \over 2 v} - \ln
\bigl(\f + {8 \over 3} \a v^2\bigr)\Bigr) - 3 \ln \Bigl(\f (1+3
v^2) + {8 \over 3} \a v^2 (1+3 v^4)\Bigr) \nonumber \\ & {  } & + 3 K_1
\end{eqnarray}
where
\begin{equation}
K_1 = - \int^\pi_\pi {d^4p \over (2 \pi)^4} \ln \bigl(1-C_1 \cos p_o -
C_2 (\cos p_1+\cos p_2 + \cos p_3)\bigr) \label{INT}
\end{equation}
with
\begin{equation}
C_1 = ( 1+3 c)^{-1}, C_2 = (3+{1 \over c})^{-1} \label{COEF}
\end{equation}
and
\begin{equation}
c = {3 \f v^2 + 8 \a v^6 \over 3 \f + 8 \a v^2}.
\end{equation}

A comparison of magnitude of $F_{tree}$ and
$F_{1-loop}$ near the phase transition line shows that they are of the
same order of magnitude, and so care should be taken in using
(\ref{FOUR}).
The loop expansion in (\ref{FOUR}) is justified by saying that the
higher loop corrections are suppressed by inverse powers of $\l$ \cite{dro}.
However, the lower loop terms might have a large contribution from higher
powers of $1 / \l$. Reexpanding $F_{tree}+F_{1-loop}$ in (\ref{FOUR})
we found that the term proportional to $1 / \l$ comes with a
coefficient $\sim O(10)$. So for consistency, one should explicitly
write the series as a series in inverse powers of $\l$. Both due to the sensitivity
of the location of the endpoint of the phase transition line, and to
check the convergence of the series in $1
/ \l$, it is necessary to evaluate the series up to the term
proportional to $1 / \l$.
This can be done \cite{fly} by a comparison with the weak coupling
perturbation series. Using the weak coupling series of Ref
\cite{das}, one gets the following expression for free energy
around this saddle point :
\begin{equation}
F_{II} = {2 + {3 r \over 2} \over 2 + 4 r} \l - {9 \over 2} \ln \l + c_0 +
{c_1 \over \l} + O\bigl({1 \over \l^2}\bigr)  \label{MOD}
\end{equation}
where $r = 4 \a / 3 \f$, and the coefficients are
\begin{eqnarray}
c_0 & = & - {3 \over 2} \ln \pi + {9 \over 2} \ln 3 - 3 \ln 2 + {(28 + 157 r + 102 r^2)
\over 8 (1 + 2 r)^2} + 3 K_2, \\
c_1 & = & 5.4 - {68 + 1115 r + 804 r^2 + 1692 r^3 \over 32 (1 + 2 r)^3} - {9 (3 + r) \over 4 (1 + 2 r)}
\end{eqnarray}
and
\begin{eqnarray}
K_2 & = & - \int^\pi_{- \pi} {d^4p \over (2 \pi)^4} \ln \Bigl(1 - {1 \over 4} \sum_{\nu=0}^3 \cos p_\nu\Bigr) \\
  & \approx & .0798.
\end{eqnarray}
It is assuring to note that the coefficient of the $O(1 / \l) $
term is $\sim O(1)$ - large contributions from $F_{1-loop}$ and
$F_{2-loop}$ cancel to give a well-behaved series. In the following,
we will use eqn(\ref{MOD}) for the free energy in this region.

Pure Yang-Mills theory has other saddle point solutions\cite{yon} : the
``fluxon'' configurations, which are topological excitations corresponding to
center of the gauge group. These are stable for $\a > {3 \over 8} \f$
\cite{bac}.
For the region of large $\f$, contributions of these maxima
are much smaller than the higher loop terms and so we will neglect these configurations.

Since the term proportional to $\a$ in (\ref{BHA}) is blind to the center of the
group,  such configurations are not
suppressed on or near the $\f = 0$ axis, and will have to be taken into account. This is done 
\cite{alb} by expanding the partition function in a series in $\f$,
to obtain 
\begin{equation}
F_{III}(\f, \a) = - 6 \f \bigl (1 - {1 \over 2} \f \bigr ) + F_{SO(3)} \bigl(\a+{3 \over 8}
\f^2\bigr) + O \bigl( \f^4 \bigr)  \label{FIVE}
\end{equation} 
where, up to one loop,
\begin{eqnarray}
F_{SO(3)} (\a) & \approx & 3 \Bigl(\a (v^2+v^4 - 2) + \ln\bigl(I_0
(\l) - I_1 (\l)\bigr) + {\l \over 2} - {3 \over 2} \l v\Bigr) \nonumber \\ &{ }& + {3 \over 2} \Bigl(\ln2+3 \ln\bigl(1+2 v^2\bigr)- 2 \ln\bigl(1+3 v^2\bigr)+ 2 K_1\Bigr).
\end{eqnarray}
Here $K_1$ is given by eqn(\ref{INT}) with a substitution of $c=v^4$
in (\ref{COEF}). $\l,v$ are solutions of
\begin{eqnarray}
v & = & {1 \over 3} {I_1(\l) - I_2(\l) \over
I_0(\l) - I_1(\l)}, \\
\l & = & {4 \over 3} \a v (1+2 v^2).
\end{eqnarray}
One can check that coefficient of the term proportional to $1 / \l$ in
$F_{SO(3)}$ is $\sim O(1)$ - the loop expansion does not differ
substantially from an expansion in $1 / \l$.
Figs. 1 and 2 display a comparison of the Monte Carlo data \cite{bha, gav3}
with the predictions for phase transition lines obtained by comparing the free
energies $F_I, F_{II}, F_{III}$in the different regions. 
\vskip6mm
\begin{center}
\setlength{\unitlength}{0.240900pt}
\ifx\plotpoint\undefined\newsavebox{\plotpoint}\fi
\sbox{\plotpoint}{\rule[-0.200pt]{0.400pt}{0.400pt}}%
\begin{picture}(1500,900)(0,0)
\font\gnuplot=cmr10 at 10pt
\gnuplot
\sbox{\plotpoint}{\rule[-0.200pt]{0.400pt}{0.400pt}}%
\put(220.0,113.0){\rule[-0.200pt]{292.934pt}{0.400pt}}
\put(220.0,113.0){\rule[-0.200pt]{0.400pt}{184.048pt}}
\put(220.0,113.0){\rule[-0.200pt]{4.818pt}{0.400pt}}
\put(198,113){\makebox(0,0)[r]{0}}
\put(1416.0,113.0){\rule[-0.200pt]{4.818pt}{0.400pt}}
\put(220.0,222.0){\rule[-0.200pt]{4.818pt}{0.400pt}}
\put(198,222){\makebox(0,0)[r]{0.5}}
\put(1416.0,222.0){\rule[-0.200pt]{4.818pt}{0.400pt}}
\put(220.0,331.0){\rule[-0.200pt]{4.818pt}{0.400pt}}
\put(198,331){\makebox(0,0)[r]{1}}
\put(1416.0,331.0){\rule[-0.200pt]{4.818pt}{0.400pt}}
\put(220.0,440.0){\rule[-0.200pt]{4.818pt}{0.400pt}}
\put(198,440){\makebox(0,0)[r]{1.5}}
\put(1416.0,440.0){\rule[-0.200pt]{4.818pt}{0.400pt}}
\put(220.0,550.0){\rule[-0.200pt]{4.818pt}{0.400pt}}
\put(198,550){\makebox(0,0)[r]{2}}
\put(1416.0,550.0){\rule[-0.200pt]{4.818pt}{0.400pt}}
\put(220.0,659.0){\rule[-0.200pt]{4.818pt}{0.400pt}}
\put(198,659){\makebox(0,0)[r]{2.5}}
\put(1416.0,659.0){\rule[-0.200pt]{4.818pt}{0.400pt}}
\put(220.0,768.0){\rule[-0.200pt]{4.818pt}{0.400pt}}
\put(198,768){\makebox(0,0)[r]{3}}
\put(1416.0,768.0){\rule[-0.200pt]{4.818pt}{0.400pt}}
\put(220.0,877.0){\rule[-0.200pt]{4.818pt}{0.400pt}}
\put(198,877){\makebox(0,0)[r]{3.5}}
\put(1416.0,877.0){\rule[-0.200pt]{4.818pt}{0.400pt}}
\put(220.0,113.0){\rule[-0.200pt]{0.400pt}{4.818pt}}
\put(220,68){\makebox(0,0){0}}
\put(220.0,857.0){\rule[-0.200pt]{0.400pt}{4.818pt}}
\put(463.0,113.0){\rule[-0.200pt]{0.400pt}{4.818pt}}
\put(463,68){\makebox(0,0){0.5}}
\put(463.0,857.0){\rule[-0.200pt]{0.400pt}{4.818pt}}
\put(706.0,113.0){\rule[-0.200pt]{0.400pt}{4.818pt}}
\put(706,68){\makebox(0,0){1}}
\put(706.0,857.0){\rule[-0.200pt]{0.400pt}{4.818pt}}
\put(950.0,113.0){\rule[-0.200pt]{0.400pt}{4.818pt}}
\put(950,68){\makebox(0,0){1.5}}
\put(950.0,857.0){\rule[-0.200pt]{0.400pt}{4.818pt}}
\put(1193.0,113.0){\rule[-0.200pt]{0.400pt}{4.818pt}}
\put(1193,68){\makebox(0,0){2}}
\put(1193.0,857.0){\rule[-0.200pt]{0.400pt}{4.818pt}}
\put(1436.0,113.0){\rule[-0.200pt]{0.400pt}{4.818pt}}
\put(1436,68){\makebox(0,0){2.5}}
\put(1436.0,857.0){\rule[-0.200pt]{0.400pt}{4.818pt}}
\put(220.0,113.0){\rule[-0.200pt]{292.934pt}{0.400pt}}
\put(1436.0,113.0){\rule[-0.200pt]{0.400pt}{184.048pt}}
\put(220.0,877.0){\rule[-0.200pt]{292.934pt}{0.400pt}}
\put(45,495){\makebox(0,0){$\beta_a$}}
\put(828,23){\makebox(0,0){$\beta_f$}}
\put(220.0,113.0){\rule[-0.200pt]{0.400pt}{184.048pt}}
\put(662,479){\makebox(0,0){$+$}}
\put(562,550){\makebox(0,0){$+$}}
\put(513,592){\makebox(0,0){$+$}}
\put(474,618){\makebox(0,0){$+$}}
\put(435,635){\makebox(0,0){$+$}}
\put(391,649){\makebox(0,0){$+$}}
\put(320,654){\makebox(0,0){$+$}}
\put(220,659){\makebox(0,0){$+$}}
\put(463,659){\makebox(0,0){$+$}}
\put(458,684){\makebox(0,0){$+$}}
\put(447,768){\makebox(0,0){$+$}}
\put(950,309){\raisebox{-.8pt}{\makebox(0,0){$\Diamond$}}}
\put(817,375){\raisebox{-.8pt}{\makebox(0,0){$\Diamond$}}}
\put(778,399){\raisebox{-.8pt}{\makebox(0,0){$\Diamond$}}}
\put(778,399){\usebox{\plotpoint}}
\multiput(778,399)(-17.086,11.784){7}{\usebox{\plotpoint}}
\multiput(662,479)(-16.924,12.016){6}{\usebox{\plotpoint}}
\multiput(562,550)(-15.759,13.508){3}{\usebox{\plotpoint}}
\multiput(513,592)(-17.270,11.513){3}{\usebox{\plotpoint}}
\multiput(474,618)(-5.378,20.047){2}{\usebox{\plotpoint}}
\put(459.38,677.10){\usebox{\plotpoint}}
\multiput(458,684)(-2.695,20.580){4}{\usebox{\plotpoint}}
\multiput(447,768)(-0.951,20.734){5}{\usebox{\plotpoint}}
\put(442,877){\usebox{\plotpoint}}
\put(474,618){\usebox{\plotpoint}}
\multiput(474,618)(-19.026,8.294){3}{\usebox{\plotpoint}}
\multiput(435,635)(-19.778,6.293){2}{\usebox{\plotpoint}}
\multiput(391,649)(-20.704,1.458){3}{\usebox{\plotpoint}}
\multiput(320,654)(-20.730,1.036){5}{\usebox{\plotpoint}}
\put(220,659){\usebox{\plotpoint}}
\multiput(480.61,864.41)(0.447,-4.704){3}{\rule{0.108pt}{3.033pt}}
\multiput(479.17,870.70)(3.000,-15.704){2}{\rule{0.400pt}{1.517pt}}
\multiput(483.59,796.97)(0.477,-4.829){7}{\rule{0.115pt}{3.620pt}}
\multiput(482.17,804.49)(5.000,-36.487){2}{\rule{0.400pt}{1.810pt}}
\multiput(488.60,758.45)(0.468,-3.113){5}{\rule{0.113pt}{2.300pt}}
\multiput(487.17,763.23)(4.000,-17.226){2}{\rule{0.400pt}{1.150pt}}
\put(483.0,812.0){\rule[-0.200pt]{0.400pt}{10.359pt}}
\multiput(492.59,716.28)(0.477,-2.380){7}{\rule{0.115pt}{1.860pt}}
\multiput(491.17,720.14)(5.000,-18.139){2}{\rule{0.400pt}{0.930pt}}
\multiput(497.59,694.61)(0.477,-2.269){7}{\rule{0.115pt}{1.780pt}}
\multiput(496.17,698.31)(5.000,-17.306){2}{\rule{0.400pt}{0.890pt}}
\put(492.0,724.0){\rule[-0.200pt]{0.400pt}{5.300pt}}
\put(502.0,659.0){\rule[-0.200pt]{0.400pt}{5.300pt}}
\put(225,686){\usebox{\plotpoint}}
\put(230,684.67){\rule{1.204pt}{0.400pt}}
\multiput(230.00,685.17)(2.500,-1.000){2}{\rule{0.602pt}{0.400pt}}
\put(225.0,686.0){\rule[-0.200pt]{1.204pt}{0.400pt}}
\multiput(269.00,683.95)(10.509,-0.447){3}{\rule{6.500pt}{0.108pt}}
\multiput(269.00,684.17)(34.509,-3.000){2}{\rule{3.250pt}{0.400pt}}
\multiput(317.00,680.94)(7.061,-0.468){5}{\rule{5.000pt}{0.113pt}}
\multiput(317.00,681.17)(38.622,-4.000){2}{\rule{2.500pt}{0.400pt}}
\multiput(366.00,676.93)(4.378,-0.482){9}{\rule{3.367pt}{0.116pt}}
\multiput(366.00,677.17)(42.012,-6.000){2}{\rule{1.683pt}{0.400pt}}
\multiput(415.00,670.93)(2.766,-0.489){15}{\rule{2.233pt}{0.118pt}}
\multiput(415.00,671.17)(43.365,-9.000){2}{\rule{1.117pt}{0.400pt}}
\multiput(463.00,661.93)(2.825,-0.489){15}{\rule{2.278pt}{0.118pt}}
\multiput(463.00,662.17)(44.272,-9.000){2}{\rule{1.139pt}{0.400pt}}
\put(235.0,685.0){\rule[-0.200pt]{8.191pt}{0.400pt}}
\put(502,659){\usebox{\plotpoint}}
\multiput(502.00,657.92)(0.544,-0.496){41}{\rule{0.536pt}{0.120pt}}
\multiput(502.00,658.17)(22.887,-22.000){2}{\rule{0.268pt}{0.400pt}}
\multiput(526.00,635.92)(0.567,-0.496){41}{\rule{0.555pt}{0.120pt}}
\multiput(526.00,636.17)(23.849,-22.000){2}{\rule{0.277pt}{0.400pt}}
\multiput(551.00,613.92)(0.660,-0.496){41}{\rule{0.627pt}{0.120pt}}
\multiput(551.00,614.17)(27.698,-22.000){2}{\rule{0.314pt}{0.400pt}}
\multiput(580.00,591.92)(0.660,-0.496){41}{\rule{0.627pt}{0.120pt}}
\multiput(580.00,592.17)(27.698,-22.000){2}{\rule{0.314pt}{0.400pt}}
\multiput(609.00,569.92)(0.765,-0.496){39}{\rule{0.710pt}{0.119pt}}
\multiput(609.00,570.17)(30.527,-21.000){2}{\rule{0.355pt}{0.400pt}}
\multiput(641.00,548.92)(0.775,-0.496){41}{\rule{0.718pt}{0.120pt}}
\multiput(641.00,549.17)(32.509,-22.000){2}{\rule{0.359pt}{0.400pt}}
\multiput(675.00,526.92)(0.822,-0.496){41}{\rule{0.755pt}{0.120pt}}
\multiput(675.00,527.17)(34.434,-22.000){2}{\rule{0.377pt}{0.400pt}}
\multiput(711.00,504.92)(0.891,-0.496){41}{\rule{0.809pt}{0.120pt}}
\multiput(711.00,505.17)(37.321,-22.000){2}{\rule{0.405pt}{0.400pt}}
\multiput(750.00,482.92)(0.891,-0.496){41}{\rule{0.809pt}{0.120pt}}
\multiput(750.00,483.17)(37.321,-22.000){2}{\rule{0.405pt}{0.400pt}}
\multiput(789.00,460.92)(1.007,-0.496){41}{\rule{0.900pt}{0.120pt}}
\multiput(789.00,461.17)(42.132,-22.000){2}{\rule{0.450pt}{0.400pt}}
\multiput(833.00,438.92)(1.128,-0.496){39}{\rule{0.995pt}{0.119pt}}
\multiput(833.00,439.17)(44.934,-21.000){2}{\rule{0.498pt}{0.400pt}}
\multiput(880.00,417.92)(1.168,-0.496){41}{\rule{1.027pt}{0.120pt}}
\multiput(880.00,418.17)(48.868,-22.000){2}{\rule{0.514pt}{0.400pt}}
\multiput(931.00,395.92)(1.284,-0.496){41}{\rule{1.118pt}{0.120pt}}
\multiput(931.00,396.17)(53.679,-22.000){2}{\rule{0.559pt}{0.400pt}}
\multiput(987.00,373.92)(1.446,-0.496){41}{\rule{1.245pt}{0.120pt}}
\multiput(987.00,374.17)(60.415,-22.000){2}{\rule{0.623pt}{0.400pt}}
\multiput(1050.00,351.92)(2.393,-0.496){41}{\rule{1.991pt}{0.120pt}}
\multiput(1050.00,352.17)(99.868,-22.000){2}{\rule{0.995pt}{0.400pt}}
\multiput(1154.00,329.92)(2.000,-0.496){41}{\rule{1.682pt}{0.120pt}}
\multiput(1154.00,330.17)(83.509,-22.000){2}{\rule{0.841pt}{0.400pt}}
\sbox{\plotpoint}{\rule[-0.500pt]{1.000pt}{1.000pt}}%
\multiput(480,877)(2.804,-20.565){2}{\usebox{\plotpoint}}
\multiput(483,855)(0.000,-20.756){2}{\usebox{\plotpoint}}
\multiput(483,812)(2.343,-20.623){2}{\usebox{\plotpoint}}
\put(490.69,753.20){\usebox{\plotpoint}}
\put(492.00,732.56){\usebox{\plotpoint}}
\put(494.70,712.11){\usebox{\plotpoint}}
\put(499.41,691.89){\usebox{\plotpoint}}
\put(502.00,671.44){\usebox{\plotpoint}}
\put(502,659){\usebox{\plotpoint}}
\put(225,691){\usebox{\plotpoint}}
\put(225.00,691.00){\usebox{\plotpoint}}
\multiput(235,691)(20.756,0.000){0}{\usebox{\plotpoint}}
\multiput(244,691)(20.689,-1.655){2}{\usebox{\plotpoint}}
\multiput(269,689)(20.595,-2.574){2}{\usebox{\plotpoint}}
\multiput(317,683)(20.756,0.000){2}{\usebox{\plotpoint}}
\multiput(366,683)(20.547,-2.935){3}{\usebox{\plotpoint}}
\multiput(415,676)(20.595,-2.574){2}{\usebox{\plotpoint}}
\multiput(463,670)(20.414,-3.750){2}{\usebox{\plotpoint}}
\put(512,661){\usebox{\plotpoint}}
\put(502,663){\usebox{\plotpoint}}
\put(502.00,663.00){\usebox{\plotpoint}}
\put(517.58,649.30){\usebox{\plotpoint}}
\multiput(531,637)(15.581,-13.712){2}{\usebox{\plotpoint}}
\multiput(556,615)(16.536,-12.544){2}{\usebox{\plotpoint}}
\put(597.64,583.41){\usebox{\plotpoint}}
\multiput(614,571)(17.659,-10.907){2}{\usebox{\plotpoint}}
\multiput(648,550)(16.536,-12.544){2}{\usebox{\plotpoint}}
\multiput(677,528)(18.564,-9.282){3}{\usebox{\plotpoint}}
\multiput(721,506)(18.564,-9.282){2}{\usebox{\plotpoint}}
\multiput(765,484)(19.170,-7.957){3}{\usebox{\plotpoint}}
\multiput(818,462)(20.055,-5.348){0}{\usebox{\plotpoint}}
\put(834.60,457.66){\usebox{\plotpoint}}
\put(854.55,452.47){\usebox{\plotpoint}}
\put(857,451){\usebox{\plotpoint}}
\end{picture}
\end{center}
\vskip3mm
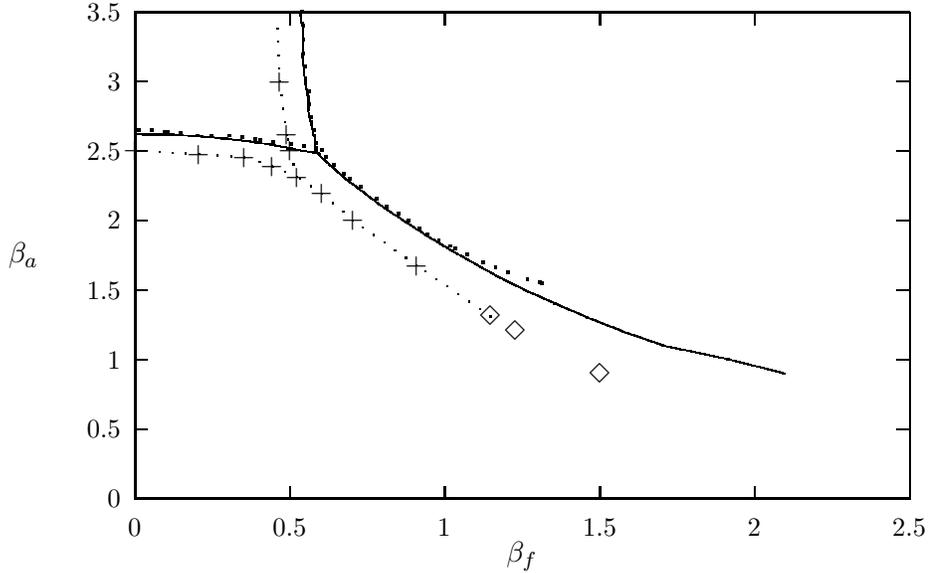
{\bf Fig.1 :} The points (joined by thin dots) are the Monte Carlo data of \cite{bha}, with
open diamonds denoting the points where a first order bulk phase transition was ruled
out by \cite{gav3}. The full line is
the curve obtained in \cite{alb}, thick dotted line is obtained taking 
upto 10th order in eqn(\ref{STRONG}).
\vskip6mm
In Fig.1 the Monte Carlo data are shown along with the curve obtained in 
\cite{alb} by comparing $F_I$ upto 10th order in eqn(\ref{STRONG})and
$F_{II}, F_{III}$ upto one loop order in eqns (\ref{FOUR}), (\ref{FIVE}).
Also shown is the curve obtained by taking the strong coupling series
upto 16th order. While
taking the higher order terms in strong coupling expansion changes the 
location of the endpoint of the phase transition line drastically,
moving it closer to the Monte Carlo data, the location of the  
transition line itself changes very little for $\a \ge 1.8$. 
\vskip6mm
\begin{center}
\setlength{\unitlength}{0.240900pt}
\ifx\plotpoint\undefined\newsavebox{\plotpoint}\fi
\sbox{\plotpoint}{\rule[-0.200pt]{0.400pt}{0.400pt}}%
\begin{picture}(1500,900)(0,0)
\font\gnuplot=cmr10 at 10pt
\gnuplot
\sbox{\plotpoint}{\rule[-0.200pt]{0.400pt}{0.400pt}}%
\put(220.0,113.0){\rule[-0.200pt]{292.934pt}{0.400pt}}
\put(220.0,113.0){\rule[-0.200pt]{0.400pt}{184.048pt}}
\put(220.0,113.0){\rule[-0.200pt]{4.818pt}{0.400pt}}
\put(198,113){\makebox(0,0)[r]{0}}
\put(1416.0,113.0){\rule[-0.200pt]{4.818pt}{0.400pt}}
\put(220.0,222.0){\rule[-0.200pt]{4.818pt}{0.400pt}}
\put(198,222){\makebox(0,0)[r]{0.5}}
\put(1416.0,222.0){\rule[-0.200pt]{4.818pt}{0.400pt}}
\put(220.0,331.0){\rule[-0.200pt]{4.818pt}{0.400pt}}
\put(198,331){\makebox(0,0)[r]{1}}
\put(1416.0,331.0){\rule[-0.200pt]{4.818pt}{0.400pt}}
\put(220.0,440.0){\rule[-0.200pt]{4.818pt}{0.400pt}}
\put(198,440){\makebox(0,0)[r]{1.5}}
\put(1416.0,440.0){\rule[-0.200pt]{4.818pt}{0.400pt}}
\put(220.0,550.0){\rule[-0.200pt]{4.818pt}{0.400pt}}
\put(198,550){\makebox(0,0)[r]{2}}
\put(1416.0,550.0){\rule[-0.200pt]{4.818pt}{0.400pt}}
\put(220.0,659.0){\rule[-0.200pt]{4.818pt}{0.400pt}}
\put(198,659){\makebox(0,0)[r]{2.5}}
\put(1416.0,659.0){\rule[-0.200pt]{4.818pt}{0.400pt}}
\put(220.0,768.0){\rule[-0.200pt]{4.818pt}{0.400pt}}
\put(198,768){\makebox(0,0)[r]{3}}
\put(1416.0,768.0){\rule[-0.200pt]{4.818pt}{0.400pt}}
\put(220.0,877.0){\rule[-0.200pt]{4.818pt}{0.400pt}}
\put(198,877){\makebox(0,0)[r]{3.5}}
\put(1416.0,877.0){\rule[-0.200pt]{4.818pt}{0.400pt}}
\put(220.0,113.0){\rule[-0.200pt]{0.400pt}{4.818pt}}
\put(220,68){\makebox(0,0){0}}
\put(220.0,857.0){\rule[-0.200pt]{0.400pt}{4.818pt}}
\put(463.0,113.0){\rule[-0.200pt]{0.400pt}{4.818pt}}
\put(463,68){\makebox(0,0){0.5}}
\put(463.0,857.0){\rule[-0.200pt]{0.400pt}{4.818pt}}
\put(706.0,113.0){\rule[-0.200pt]{0.400pt}{4.818pt}}
\put(706,68){\makebox(0,0){1}}
\put(706.0,857.0){\rule[-0.200pt]{0.400pt}{4.818pt}}
\put(950.0,113.0){\rule[-0.200pt]{0.400pt}{4.818pt}}
\put(950,68){\makebox(0,0){1.5}}
\put(950.0,857.0){\rule[-0.200pt]{0.400pt}{4.818pt}}
\put(1193.0,113.0){\rule[-0.200pt]{0.400pt}{4.818pt}}
\put(1193,68){\makebox(0,0){2}}
\put(1193.0,857.0){\rule[-0.200pt]{0.400pt}{4.818pt}}
\put(1436.0,113.0){\rule[-0.200pt]{0.400pt}{4.818pt}}
\put(1436,68){\makebox(0,0){2.5}}
\put(1436.0,857.0){\rule[-0.200pt]{0.400pt}{4.818pt}}
\put(220.0,113.0){\rule[-0.200pt]{292.934pt}{0.400pt}}
\put(1436.0,113.0){\rule[-0.200pt]{0.400pt}{184.048pt}}
\put(220.0,877.0){\rule[-0.200pt]{292.934pt}{0.400pt}}
\put(45,495){\makebox(0,0){$\beta_a$}}
\put(828,23){\makebox(0,0){$\beta_f$}}
\put(220.0,113.0){\rule[-0.200pt]{0.400pt}{184.048pt}}
\put(662,479){\makebox(0,0){$+$}}
\put(562,550){\makebox(0,0){$+$}}
\put(513,592){\makebox(0,0){$+$}}
\put(474,618){\makebox(0,0){$+$}}
\put(435,635){\makebox(0,0){$+$}}
\put(391,649){\makebox(0,0){$+$}}
\put(320,654){\makebox(0,0){$+$}}
\put(220,659){\makebox(0,0){$+$}}
\put(463,659){\makebox(0,0){$+$}}
\put(458,684){\makebox(0,0){$+$}}
\put(447,768){\makebox(0,0){$+$}}
\put(950,309){\raisebox{-.8pt}{\makebox(0,0){$\Diamond$}}}
\put(817,375){\raisebox{-.8pt}{\makebox(0,0){$\Diamond$}}}
\put(778,399){\raisebox{-.8pt}{\makebox(0,0){$\Diamond$}}}
\put(778,399){\usebox{\plotpoint}}
\multiput(778,399)(-17.086,11.784){7}{\usebox{\plotpoint}}
\multiput(662,479)(-16.924,12.016){6}{\usebox{\plotpoint}}
\multiput(562,550)(-15.759,13.508){3}{\usebox{\plotpoint}}
\multiput(513,592)(-17.270,11.513){3}{\usebox{\plotpoint}}
\multiput(474,618)(-5.378,20.047){2}{\usebox{\plotpoint}}
\put(459.38,677.10){\usebox{\plotpoint}}
\multiput(458,684)(-2.695,20.580){4}{\usebox{\plotpoint}}
\multiput(447,768)(-0.951,20.734){5}{\usebox{\plotpoint}}
\put(442,877){\usebox{\plotpoint}}
\put(474,618){\usebox{\plotpoint}}
\multiput(474,618)(-19.026,8.294){3}{\usebox{\plotpoint}}
\multiput(435,635)(-19.778,6.293){2}{\usebox{\plotpoint}}
\multiput(391,649)(-20.704,1.458){3}{\usebox{\plotpoint}}
\multiput(320,654)(-20.730,1.036){5}{\usebox{\plotpoint}}
\put(220,659){\usebox{\plotpoint}}
\put(452,877){\usebox{\plotpoint}}
\put(451.67,833){\rule{0.400pt}{5.300pt}}
\multiput(451.17,844.00)(1.000,-11.000){2}{\rule{0.400pt}{2.650pt}}
\put(452.67,812){\rule{0.400pt}{5.059pt}}
\multiput(452.17,822.50)(1.000,-10.500){2}{\rule{0.400pt}{2.529pt}}
\put(452.0,855.0){\rule[-0.200pt]{0.400pt}{5.300pt}}
\put(453.67,768){\rule{0.400pt}{5.300pt}}
\multiput(453.17,779.00)(1.000,-11.000){2}{\rule{0.400pt}{2.650pt}}
\put(454.67,746){\rule{0.400pt}{5.300pt}}
\multiput(454.17,757.00)(1.000,-11.000){2}{\rule{0.400pt}{2.650pt}}
\put(456.17,724){\rule{0.400pt}{4.500pt}}
\multiput(455.17,736.66)(2.000,-12.660){2}{\rule{0.400pt}{2.250pt}}
\put(457.67,702){\rule{0.400pt}{5.300pt}}
\multiput(457.17,713.00)(1.000,-11.000){2}{\rule{0.400pt}{2.650pt}}
\put(458.67,681){\rule{0.400pt}{5.059pt}}
\multiput(458.17,691.50)(1.000,-10.500){2}{\rule{0.400pt}{2.529pt}}
\put(459.67,667){\rule{0.400pt}{3.373pt}}
\multiput(459.17,674.00)(1.000,-7.000){2}{\rule{0.400pt}{1.686pt}}
\put(454.0,790.0){\rule[-0.200pt]{0.400pt}{5.300pt}}
\put(225,688){\usebox{\plotpoint}}
\put(244,686.67){\rule{6.023pt}{0.400pt}}
\multiput(244.00,687.17)(12.500,-1.000){2}{\rule{3.011pt}{0.400pt}}
\put(269,685.67){\rule{5.782pt}{0.400pt}}
\multiput(269.00,686.17)(12.000,-1.000){2}{\rule{2.891pt}{0.400pt}}
\put(293,684.67){\rule{5.782pt}{0.400pt}}
\multiput(293.00,685.17)(12.000,-1.000){2}{\rule{2.891pt}{0.400pt}}
\put(317,683.17){\rule{5.100pt}{0.400pt}}
\multiput(317.00,684.17)(14.415,-2.000){2}{\rule{2.550pt}{0.400pt}}
\put(342,681.17){\rule{4.900pt}{0.400pt}}
\multiput(342.00,682.17)(13.830,-2.000){2}{\rule{2.450pt}{0.400pt}}
\put(366,679.17){\rule{4.900pt}{0.400pt}}
\multiput(366.00,680.17)(13.830,-2.000){2}{\rule{2.450pt}{0.400pt}}
\multiput(390.00,677.95)(5.374,-0.447){3}{\rule{3.433pt}{0.108pt}}
\multiput(390.00,678.17)(17.874,-3.000){2}{\rule{1.717pt}{0.400pt}}
\multiput(415.00,674.94)(3.406,-0.468){5}{\rule{2.500pt}{0.113pt}}
\multiput(415.00,675.17)(18.811,-4.000){2}{\rule{1.250pt}{0.400pt}}
\multiput(439.00,670.95)(5.151,-0.447){3}{\rule{3.300pt}{0.108pt}}
\multiput(439.00,671.17)(17.151,-3.000){2}{\rule{1.650pt}{0.400pt}}
\put(225.0,688.0){\rule[-0.200pt]{4.577pt}{0.400pt}}
\put(463,667){\usebox{\plotpoint}}
\multiput(463.59,664.69)(0.485,-0.569){11}{\rule{0.117pt}{0.557pt}}
\multiput(462.17,665.84)(7.000,-6.844){2}{\rule{0.400pt}{0.279pt}}
\multiput(470.58,656.76)(0.496,-0.549){37}{\rule{0.119pt}{0.540pt}}
\multiput(469.17,657.88)(20.000,-20.879){2}{\rule{0.400pt}{0.270pt}}
\multiput(490.00,635.92)(0.521,-0.496){41}{\rule{0.518pt}{0.120pt}}
\multiput(490.00,636.17)(21.924,-22.000){2}{\rule{0.259pt}{0.400pt}}
\multiput(513.00,613.92)(0.544,-0.496){41}{\rule{0.536pt}{0.120pt}}
\multiput(513.00,614.17)(22.887,-22.000){2}{\rule{0.268pt}{0.400pt}}
\multiput(537.00,591.92)(0.591,-0.496){41}{\rule{0.573pt}{0.120pt}}
\multiput(537.00,592.17)(24.811,-22.000){2}{\rule{0.286pt}{0.400pt}}
\multiput(563.00,569.92)(0.668,-0.496){39}{\rule{0.633pt}{0.119pt}}
\multiput(563.00,570.17)(26.685,-21.000){2}{\rule{0.317pt}{0.400pt}}
\multiput(591.00,548.92)(0.660,-0.496){41}{\rule{0.627pt}{0.120pt}}
\multiput(591.00,549.17)(27.698,-22.000){2}{\rule{0.314pt}{0.400pt}}
\multiput(620.00,526.92)(0.706,-0.496){41}{\rule{0.664pt}{0.120pt}}
\multiput(620.00,527.17)(29.623,-22.000){2}{\rule{0.332pt}{0.400pt}}
\multiput(651.00,504.92)(0.752,-0.496){41}{\rule{0.700pt}{0.120pt}}
\multiput(651.00,505.17)(31.547,-22.000){2}{\rule{0.350pt}{0.400pt}}
\multiput(684.00,482.92)(0.775,-0.496){41}{\rule{0.718pt}{0.120pt}}
\multiput(684.00,483.17)(32.509,-22.000){2}{\rule{0.359pt}{0.400pt}}
\multiput(718.00,460.92)(0.799,-0.496){41}{\rule{0.736pt}{0.120pt}}
\multiput(718.00,461.17)(33.472,-22.000){2}{\rule{0.368pt}{0.400pt}}
\multiput(753.00,438.92)(0.910,-0.496){39}{\rule{0.824pt}{0.119pt}}
\multiput(753.00,439.17)(36.290,-21.000){2}{\rule{0.412pt}{0.400pt}}
\multiput(791.00,417.92)(0.845,-0.496){41}{\rule{0.773pt}{0.120pt}}
\multiput(791.00,418.17)(35.396,-22.000){2}{\rule{0.386pt}{0.400pt}}
\multiput(828.00,395.92)(0.845,-0.496){41}{\rule{0.773pt}{0.120pt}}
\multiput(828.00,396.17)(35.396,-22.000){2}{\rule{0.386pt}{0.400pt}}
\multiput(865.00,373.92)(0.822,-0.496){41}{\rule{0.755pt}{0.120pt}}
\multiput(865.00,374.17)(34.434,-22.000){2}{\rule{0.377pt}{0.400pt}}
\multiput(901.00,351.92)(0.845,-0.496){41}{\rule{0.773pt}{0.120pt}}
\multiput(901.00,352.17)(35.396,-22.000){2}{\rule{0.386pt}{0.400pt}}
\multiput(938.00,329.92)(0.937,-0.496){41}{\rule{0.845pt}{0.120pt}}
\multiput(938.00,330.17)(39.245,-22.000){2}{\rule{0.423pt}{0.400pt}}
\multiput(979.00,307.92)(1.381,-0.491){17}{\rule{1.180pt}{0.118pt}}
\multiput(979.00,308.17)(24.551,-10.000){2}{\rule{0.590pt}{0.400pt}}
\sbox{\plotpoint}{\rule[-0.500pt]{1.000pt}{1.000pt}}%
\put(452,877){\usebox{\plotpoint}}
\multiput(452,877)(0.000,-20.756){2}{\usebox{\plotpoint}}
\put(452.89,835.51){\usebox{\plotpoint}}
\put(453.87,814.78){\usebox{\plotpoint}}
\put(454.00,794.02){\usebox{\plotpoint}}
\put(454.76,773.29){\usebox{\plotpoint}}
\put(455.70,752.55){\usebox{\plotpoint}}
\put(457.29,731.86){\usebox{\plotpoint}}
\put(458.58,711.15){\usebox{\plotpoint}}
\put(459.55,690.42){\usebox{\plotpoint}}
\put(461.03,669.72){\usebox{\plotpoint}}
\put(462,659){\usebox{\plotpoint}}
\put(220,687){\usebox{\plotpoint}}
\put(220.00,687.00){\usebox{\plotpoint}}
\put(240.76,687.00){\usebox{\plotpoint}}
\put(261.51,687.00){\usebox{\plotpoint}}
\put(282.22,685.90){\usebox{\plotpoint}}
\put(302.93,684.59){\usebox{\plotpoint}}
\put(323.65,683.47){\usebox{\plotpoint}}
\multiput(342,682)(20.684,-1.724){2}{\usebox{\plotpoint}}
\put(385.62,677.55){\usebox{\plotpoint}}
\put(406.23,675.05){\usebox{\plotpoint}}
\put(426.76,672.04){\usebox{\plotpoint}}
\put(447.23,668.63){\usebox{\plotpoint}}
\put(463,666){\usebox{\plotpoint}}
\put(461,667){\usebox{\plotpoint}}
\put(461.00,667.00){\usebox{\plotpoint}}
\multiput(469,659)(14.331,-15.014){2}{\usebox{\plotpoint}}
\put(504.52,622.48){\usebox{\plotpoint}}
\multiput(512,615)(16.090,-13.111){2}{\usebox{\plotpoint}}
\put(551.43,581.60){\usebox{\plotpoint}}
\multiput(563,571)(16.383,-12.743){2}{\usebox{\plotpoint}}
\multiput(590,550)(16.737,-12.274){3}{\usebox{\plotpoint}}
\multiput(650,506)(17.188,-11.635){4}{\usebox{\plotpoint}}
\multiput(715,462)(17.753,-10.752){4}{\usebox{\plotpoint}}
\multiput(786,419)(17.710,-10.823){4}{\usebox{\plotpoint}}
\multiput(858,375)(17.500,-11.160){4}{\usebox{\plotpoint}}
\multiput(927,331)(17.055,-11.829){4}{\usebox{\plotpoint}}
\multiput(989,288)(16.638,-12.408){3}{\usebox{\plotpoint}}
\multiput(1048,244)(17.840,-10.608){5}{\usebox{\plotpoint}}
\put(1139.39,193.62){\usebox{\plotpoint}}
\put(1152,189){\usebox{\plotpoint}}
\end{picture}
\end{center}
\vskip3mm
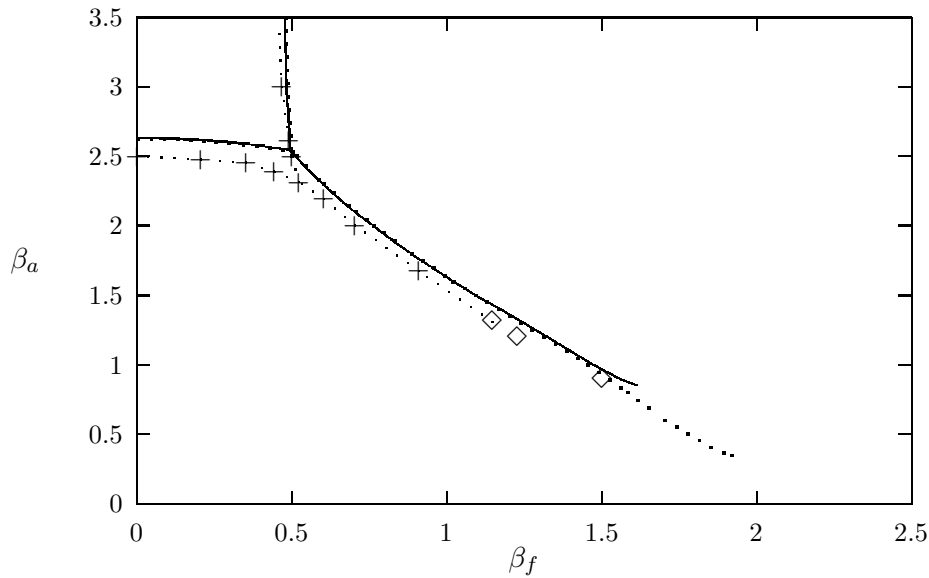
{\bf Fig.2 :} The Monte Carlo data are shown against prediction 
using 16th order strong coupling expansion and eqns (\ref{MOD}),(\ref{FIVE}) (full line). 
The thick dotted line corresponds to 10th order in strong coupling series.
\vskip6mm
In Fig.2 we show the prediction for phase transition curve
obtained by comparing the improved mean field series of eqn(\ref{MOD}),
the strong coupling series upto 16th order and eqn(\ref{FIVE}). Also shown is the
curve for taking strong coupling series upto 10th order. The same
feature, namely, the extreme sensitivity of the endpoint and robustness of 
the upper part of the curve is noted. Also, a comparison of Figs. 1 and 2
reveals that the improvement in the  
mean field series leads to a 
curve that is considerably closer to the Monte Carlo results. The
convergence of the mean field series was checked by checking that on
taking just the first two terms in (\ref{MOD}), the transition line is
left unchanged. The endpoint of the transition line,
however, is still in disagreement with \cite{gav3}. Due to the extreme
sensitivity of the endpoint, it seems that very high orders in strong
coupling series may be needed in order to pinpoint it precisely.
However, since the strong coupling expansion seems to be well behaved 
near the phase transition line for $\a \ge 1.5$, the existence of the 
bulk transition at least upto this region seems to be confirmed. 

The only caveat for this conclusion is the choice of gauge fixing. It
is always advisable to check for possible gauge dependence of a result
obtained by gauge fixing. In this case, it is even more so since the
axial gauge constrains the deconfinement order parameter to be nonzero
in region II whereas it is zero (or small) in the strong coupling
region I. It would be interesting to confirm the bulk phase diagram in
Fig.2 by using another gauge condition to establish the bulk
transition for $\a \sim 1.5$ beyond doubt.


\begin{thebibliography}{99}
 
\bibitem{wil} K. Wilson, {\em Phys. Rev.}
{\bf D10}, 2445 (1970).

\bibitem{bha} G. Bhanot and M. Creutz, {\em Phys. Rev.}
{\bf D24}, 3212 (1981).

\bibitem{gav1} R.V. Gavai, M. Grady and M. Mathur, {\em Nucl. Phys.}
{\bf B423}, 123 (1994).

\bibitem{mat} M. Mathur and R.V. Gavai, {\em Nucl. Phys.}
{\bf B448}, 399 (1995).

\bibitem{gav2} R.V. Gavai and M. Mathur, hep-lat/9512015.

\bibitem{gav3} R.V. Gavai, {\em Nucl. Phys.}
{\bf B474}, 446.

\bibitem{bor} C. Borgs and E. Seiler, {\em Nucl. Phys.}
{\bf B215}, 125 (1983).

\bibitem{alb} J.M. Alberty, H. Flyvbjerg and B. Lautrup, {\em Nucl. Phys.}
{\bf B220}, 61 (1983).

\bibitem{bre} E. Brezin and J.M. Drouffe, {\em Nucl. Phys.}
{\bf B200}, 93 (1982).

\bibitem{dro} J.M. Drouffe and J.B. Zuber, {\em Phys. Rep.}
{\bf 102}, 1 (1983).

\bibitem{das} R. Dashen, U.M. Heller and H. Neuberger, {\em Nucl. Phys.}
{\bf B215}, 360 (1983).

\bibitem{yon} T. Yoneya, {\em Nucl. Phys.}
{\bf B144}, 195 (1978).

\bibitem{bac} C.P. Bachas and R.F. Dashen, {\em Nucl. Phys.}
{\bf B210}, 583 (1982)

\bibitem{fly} H. Flyvbjerg, {\em Nucl. Phys.}
{\bf B235}, 331 (1984).


\end{thebibliography}
\end{document}